\documentclass[acmsmall]{acmart}
\usepackage{makecell} 
\usepackage{algorithm}
\usepackage{algpseudocode}
\usepackage{multirow}

\AtBeginDocument{%
  \providecommand\BibTeX{{%
    \normalfont B\kern-0.5em{\scshape i\kern-0.25em b}\kern-0.8em\TeX}}}

\setcopyright{acmcopyright}
\copyrightyear{2023} 
\acmYear{2023} 
\acmBooktitle{ACM Transactions on Information Systems}

\usepackage{subcaption}
\usepackage{graphics}
\begin{document}
\title{Generalized Weak Supervision for Neural Information Retrieval}


\author{Yen-Chieh Lien}
\affiliation{%
  \institution{University of Massachusetts Amherst, Amherst, MA}
}
\email{ylien@cs.umass.edu}


\author{Hamed Zamani}
\affiliation{%
  \institution{University of Massachusetts Amherst, Amherst, MA}
}
\email{zamani@cs.umass.edu}

\author{W. Bruce Croft}
\affiliation{%
  \institution{University of Massachusetts Amherst, Amherst, MA}
}
\email{croft@cs.umass.edu}

\renewcommand{\shortauthors}{Y.-C. Lien et al.}

\begin{abstract}

Neural ranking models (NRMs) have demonstrated effective performance in several information retrieval (IR) tasks. However, training NRMs often requires large-scale training data, which is difficult and expensive to obtain. To address this issue, one can train NRMs via weak supervision, where a large dataset is automatically generated using an existing ranking model (called the weak labeler) for training NRMs. Weakly supervised NRMs can generalize from the observed data and significantly outperform the weak labeler. This paper generalizes this idea through an iterative re-labeling process, demonstrating that weakly supervised models can iteratively play the role of weak labeler and significantly improve ranking performance without using manually labeled data. The proposed Generalized Weak Supervision (GWS) solution is generic and orthogonal to the ranking model architecture. This paper offers four implementations of GWS: self-labeling, cross-labeling, joint cross- and self-labeling, and greedy multi-labeling. GWS also benefits from a query importance weighting mechanism based on query performance prediction methods to reduce noise in the generated training data. We further  draw a theoretical connection between self-labeling and Expectation-Maximization. Our experiments on two passage retrieval benchmarks suggest that all implementations of GWS lead to substantial improvements compared to weak supervision in all cases. 



\end{abstract}



\keywords{weak supervision, neural ranking models, zero-shot learning, unsupervised learning.}


\maketitle

\section{Introduction}
\label{sec:intro}
Deep neural networks have shown promising results in many retrieval tasks, including ad-hoc retrieval \cite{DUET,DRMM,bertrank1,KD1}, conversational search \cite{conver1,conver2,convers3}, and cross-modal retrieval \cite{cross1,cross2}. Training existing neural ranking models (NRMs) often require large amount of training data. However, obtaining such a large training set is often difficult and expensive.

This paper focuses on training NRMs when no manually labeled data is available for training. A straightforward solution to tackle this problem is to use large-scale pre-trained language models, e.g., BERT \cite{BERT}, as zero-shot ranking models. However, since these models are not optimized for retrieval tasks, their zero-shot performance for retrieval tasks is limited. They even perform poorer than term matching models, such as BM25 \cite{bm25trec}. This is why these models are often fine-tuned using labeled training data.

An alternative solution to this problem is to train NRMs using noisy training signals produced by existing (unsupervised) retrieval models. This teacher-student learning approach is called \emph{weak supervision}  \cite{WS,theoryWS} and the teacher model is often called the weak labeler. 
Weak supervision addresses the data scarcity issue by leveraging unsupervised methods to infer a noisy ranked list and uses that signal as ground truth for training a neural ranking model. In this line of research, classical IR methods, such as BM25, are usually selected as the weak labeler \cite{WS,rerank}.  
There exist numerous theoretical and empirical evidence that weakly supervised models can significantly outperform their weak labelers \cite{WS,rerank,theoryWS}. This paper generalizes the weak supervision formulation such that a weakly supervised model iteratively becomes the weak labeler. We hypothesize that such an approach should lead to performance improvement since the quality of weakly supervised training data is iteratively improved. Based on this hypothesis, we propose \emph{Generalized Weak Supervision (GWS)}, a generic framework for training neural ranking models with no labeled data. We offer four implementations of this framework. The first implementation is called \emph{self-labeling}, in which one weakly supervised model iteratively produces the training data for the next iteration. The second implementation is called \emph{cross-labeling}. In this case, we use two NRMs $M_1$ and $M_2$ exchanging information by playing the roles of teacher models for one another in weak supervision. In other words, each model is optimized using the training data produced by the other model and the role of teacher model alternates between them. The third implementation is called \emph{joint cross- and self-labeling}. As the combination of the previous two implementation, we also have two NRMs $M_1$ and $M_2$ as teacher and student models. Different from \emph{cross-labeling}, which exchanges weak signals in each iteration, after each model alternation (i.e., cross-labeling), we apply \emph{self-labeling} to train the student model thoroughly and then repeat the cross-labeling process. The last implementation is called \emph{greedy multi-labeling}, in which we train several model checkpoints based on weak supervision signals generated from all ranking models and pick the best one to represent this structure as the signal provider (i.e., the teacher) for the next iteration. In other words, the best performing students at every iteration become the teacher for the next iteration.


This paper also draws theoretical connections between the simplest implementation of the proposed GWS framework (i.e., self-labeling) and the Expectation–maximization (EM) algorithm, a well-known framework for unsupervised learning which has been successfully used for a wide range of tasks including semi-supervised text classification \cite{textem}, transfer learning \cite{trans}, language model estimation \cite{Hiemstra:2004:Parsimonious}, and pseudo-relevance feedback \cite{Zhai:2001:MixtureModel}. 


We further survey techniques for enhancing the effectiveness of GWS training. To this aim, we study query importance for the weak supervision training process. Intuitively, we would like to train NRMs by emphasizing on the queries for which the weak labeler produces high quality results. This will reduce the level of noise in the training set. Based on this intuition, we leverage existing query performance prediction (QPP) models that have been studied in the information retrieval literature for decades~\cite{Cronen-Townsend:2002:Clarity,NQPP,NQC}, and propose an in-batch weighting method of training instances to modify the importance of queries based on the prediction of QPP models.

In our experiments, we evaluate the proposed methods using two publicly available datasets: (1) WikiPassageQA \cite{wikipassqa}, a passage retrieval dataset based on Wikipedia articles; and (2) ANTIQUE \cite{antique} a passage retrieval dataset for non-factoid questions submitted by real users to community question answering websites.
%
In our experiments, we follow \citet{WS} and adopt BM25 \cite{bm25} as the initial weak labeler. We train two pre-trained language models, BERT \cite{BERT} and RoBERTa \cite{roberta}, using (generalized) weak supervision. For QPP, we use Normalized Query Commitment (NQC) \cite{NQC}, a popular unsupervised QPP method for predicting the performance of the initial weak labeler for each training query. Further, we adopt an in-batch re-weighting through training process to incorporate NQC into the loss function.

To summarize, the contributions of this paper include:
\begin{itemize}
    \item Proposing the Generalized Weak Supervision framework.
    \item Introducing four implementations of the GWS framework.
    \item Drawing connections between GWS and the EM algorithm.
    \item Enhancing GWS training by leveraging query performance prediction models for query importance.
    \item Demonstrating substantial improvements in ranking quality compared to zero-shot and weakly supervised baselines. For instance, on WikiPassageQA, our best performing BERT model achieves 42.7\% and 20.8\% relative NDCG@10 improvements compared to BM25 (the initial weak labeler) and weakly supervised BERT, respectively.
    Note that these improvements are solely observed using automatic relabeling of training data with no manually labeled data.
\end{itemize}

\section{Related Work}
In this section, we first review two of the most relevant lines of research to this paper: neural ranking models and weak supervision. We further provide a brief review of prior work on self-labeling, knowledge distillation, and domain adaptation. Even though these three topics are not directly related to the contributions of this work, there are some connections that are worth exploring.

\subsection{Neural Ranking Model}
In recent years, several neural ranking models were proposed for retrieval tasks. DSSM \cite{DSSM} and C-DSSM \cite{CDSSM} adopted a method to learn the representation of query and document individually and use a matching function to score. The deep relevance matching model \cite{DRMM} exploits histogram feature to represent the interaction between query and document as the input of neural ranking architecture. DUET \cite{DUET} uses two networks to learn local interaction and distributed matching between query and document respectively. 

After BERT \cite{BERT} is proposed, large-scale pre-trained language models are widely applied to ranking problems. For instance, \citet{bertrank1} used BERT for passage ranking and demonstrated significant improvement. \citet{bertrank2} combined learning-to-rank and the ensemble of BERT \cite{BERT}, RoBERTa \cite{roberta} and ELECTRA \cite{elec} for passage ranking. \citet{bertQA} apply BERT for the conversational question answering task.

The mentioned neural ranking models focus on re-ranking problems, where an efficient first-stage retrieval model, such as BM25, provide a small list of documents for re-ranking. \citet{rerank} demonstrated for the first time that neural models can be used for document retrieval from a large collection without the need to a multi-stage cascaded retrieval architecture. This phenomena was later adopted and applied to dense query and document embedding with the use of approximate nearest neighbor search algorithms. Such dense retrieval approaches, such as DPR \cite{DPR} use a dual encoder architecture to encode queries and documents separately and compute their similarity using simple matching functions, such as dot product or cosine similarity. Several works \cite{RocketQAv2, colbert, ance, KD1} were proposed based on the dense retrieval setting to move from re-ranking to ranking. 

The vast majority of recent neural ranking models are trained on large data collections, such as MS MARCO, and do not focus on the issue of data volume. In this work, we aim to propose a general framework for training neural ranking models without a need to ground truth labels. Therefore, the proposed approach can potentially be applied to any of the existing neural ranking model architectures listed above.

\subsection{Weak Supervision for IR}
As the motivation of this paper, weak supervision try to solve the problem of data volume for neural ranking models. \citet{WS} first proposed weak supervision to train a neural ranking model based on the labels generated by existing retrieval methods, e.g., BM25, or heuristics. 
The empirically showed that weakly supervised neural ranking models can significantly improve their weak labeler, solving an important problem in optimizing large deep learning models without labeled data. Later, \citet{theoryWS} provided theoretical insights into weak supervision for information retrieval.

Several works \cite{weak2,NQPP,weak3} exploited weak supervision on specific IR tasks. \citet{weak2} used automatically generated data for fact ranking in a knowledge graph. \citet{NQPP} leveraged multiple weak signals for query performance prediction (QPP). \citet{weak3} used weak supervision to train the retrieval model with multi-level matching. \citet{RWE} used weak supervision for learning relevance-based word embeddings. 

Given the success of weak supervision in IR, a number of approaches focused on strengthening the effectiveness of weak supervision. \citet{select} applied reinforcement learning to select anchor-document pairs for training weakly supervise neural ranking models. Some methods \cite{adhoc-bert,bert-pass} adopt pre-trained language models like BERT as the weakly supervised ranking model. In this paper, we also follow this setting and use pre-train language models as the retrieval model. 

Previous work solely rely on one or more weak labeler to train their model. In this paper, we generalize this approach such that the weakly supervised models in each step become weak labelers in the next step. The proposed framework is sufficiently generatic to be applied to any weakly supervised model.


\subsection{Self-Labeling}
Self-labeling is widely used for semi-supervised learning problems. By directly imputing ground truth labels for unlabeled instance, self-labeling propagate labels to unknown target data. \citet{textem} applied self-labeling to semi-supervised text classification using an Expectation-Maximization (EM) algorithm. \citet{coda} designed an algorithm for semi-supervised sentiment classification by iterative imputing sentiment labels for unlabeled reviews according to the current model’s confidence score on the data.

Among all, we found the one conducted by \citet{trans} the most relevant to ours. The authors trained an initial ranker from ground truth labels on a source domain and used self-labeling to label the target domain's data. Then, they re-trained the ranker on the target data from self-labeling and repeated the above operation until convergence. The steps for the target domain are similar to ours, however, their task is transfer learning, which needs large-scale ground truth labels on the source domain. In our setting, we do not include any labeled data at any stage of our training.

\subsection{Knowledge Distillation for IR}
To achieve the performance of neural models with lower computational cost, a common approach is to distill knowledge from large teacher neural models into smaller student models. For pre-trained language model like BERT, DistilBERT and Tinybert are proposed to create light-weight model when maintaining the performance on various tasks using distillation.

Due to the success of pre-trained language models on IR tasks, there are several works on applying knowledge distillation on IR. \citet{KD1} proposed a curriculum learning framework to optimize student dense retrieval models from teacher re-ranking models. \citet{KD2} proposed a new cross-encoder architecture to transfer its knowledge to a low-cost bi-encoder for the response retrieval task. 
\citet{KD3} proposed a cross-architecture training procedure to adapt knowledge distillation to the varying output score distributions from different neural models.

Although the relationship between teacher and student models is similar to weak signals and weakly supervised models, there are two main differences between the two tasks. First, in the setting of knowledge distillation, label information is available especially for the teacher models' training, and the goal is to create low-cost inference when having a supervised model. Second, knowledge distillation use a smaller student model to approximate the performance of a larger one, while this is not the case in weak supervision and some of labels can even come from simple non-ML models.

\subsection{Domain Adaptation for Neural IR}
Because there exist some massive IR datasets like MS MARCO as a rich source domain, domain adaption is also a crucial solution for neural IR to solve the high dependency on in-domain data.  \citet{domain1} did early work on domain adaptation for neural retrieval by cross-domain adversarial learning, but it had not included pre-trained models from a source domain.

Recent works exploited pre-trained IR models from existing data on other retrieval tasks. \citet{GPL} trained doc2query T5 model and retrieval models on the source domain, used T5 to generate pseudo-queries on the target domain and then apply a pre-trained dense retrieval model and a cross-encoder model to build pseudo pairwise data for training a new retrieval model on the target domain. \citet{MARank} not only built pseudo-labeled data on the target domain by pre-train models on the source domain but also added a meta-learning method to learn meta weighting on synthetic data to exploit weak supervision signals better. Different from weak supervision, \citet{disent} split a retrieval component into two modules, Relevance Estimation Module (REM) and Domain Adaption Module (DAM), to deal with general relevance matching and adaptation to the target domains. Even though domain adaption focuses on solving data scarcity and sometimes include weak supervision, we deal with a different problem, which does not assume the existence of a rich source domain with large-scale labeled data. However, for the works adopting weak supervision as a part of the solution, the proposed GWS can potentially be incorporated.

\section{Formulating Weak Supervision for IR}
Given a query $q$ and a document collection $C$, the task of ad-hoc information retrieval is to develop a retrieval model $M_\theta$ parameterized by $\theta$ for retrieving documents from $C$ with respect to their relevance to the query $q$ in descending order. Unsupervised approaches for ad-hoc retrieval mostly focus on term matching between the query and document content, such as TF-IDF \cite{IDF1} , BM25 \cite{bm25trec}, and query likelihood \cite{QL}. There also exist supervised ranking models that learn from a manually labeled training set. Weak supervision is an approach for training retrieval models without any manually labeled data. It uses an unsupervised retrieval models (called the weak labeler), e.g., BM25, to automatically annotate queries and documents for training learning to rank models. 

For every training query $q \in Q$, weak supervision uses a weak labeler $\widehat{M}$ to retrieve a list of documents $D$ from $C$ and creates a set of triplets $T_{\widehat{M}} = \{(q, d, \widehat{M}(q,d)) : \forall q \in Q, \forall d \in D\}$. This training set can be considered as noisy ground truth and thus can be used for training weak supervision models as follows: 
\begin{align}
    \theta^* &= \arg \min_\theta \mathcal{L}(M_\theta, T_{\widehat{M}}) \nonumber 
\end{align}
where $Q$ and $\mathcal{L}$ denote the training query set and the ranking loss function, respectively. 
The query set $Q$ can be sampled from a search engine's query logs or questions in community question answering forums. It can  also be automatically generated using autoregressive query generation models or even by random n-gram selection from a corpus. The loss function $ \mathcal{L}$ can be implemented using any of the pointwise, pairwise, and listwise ranking loss functions. Zamani and Croft~\cite{theoryWS} proved that the weak supervision loss function $ \mathcal{L}$ should be symmetric in order to be robust to the weak supervision noise. They demonstrated that Hinge loss satisfies this property.

\section{Generalized Weak Supervision}
In this section, we introduce the generalized weak supervision (GWS) framework for information retrieval. GWS is a general framework for training ranking models. The model parameters in GWS are first initialized using typical weak supervision approaches. Next, GWS runs an iterative process. In each iteration, it re-labels the training data and uses the new training set for training another ranking model. GWS repeats this process until a stopping criterion is met. 

GWS can work with one single ranking model or multiple ranking models by changing the re-labeling settings. In this work, we provide four different settings.
\begin{enumerate}
    \item Algorithm~\ref{alg:self-labeling} introduces GWS with self-labeling, in which a single ranking model iteratively re-labels the dataset and reuses it for optimization.
    \item Algorithm~\ref{alg:cross-labeling}, on the other hand, introduces the weak labeling alternation implementation of GWS, in which the relabeling process alternates between $k$ weakly supervised rankers.
    \item Algorithm \ref{alg:CLSL} is the combination of the above two. Ranking models also provide weak signals to the other model in the manner of weak supervision but apply Algorithm~\ref{alg:self-labeling} to train a model thoroughly before exchanging.
    \item Finally, Algorithm \ref{alg:greedy-cross} also adopt a multi-model setting, but it considers all teacher-student combinations in each iterations and chooses the best one for a model structure as the teacher model for the next round.
\end{enumerate}

\begin{figure}[t]
\begin{subfigure}{.5\textwidth}
  \centering
  \includegraphics[width=.8\linewidth]{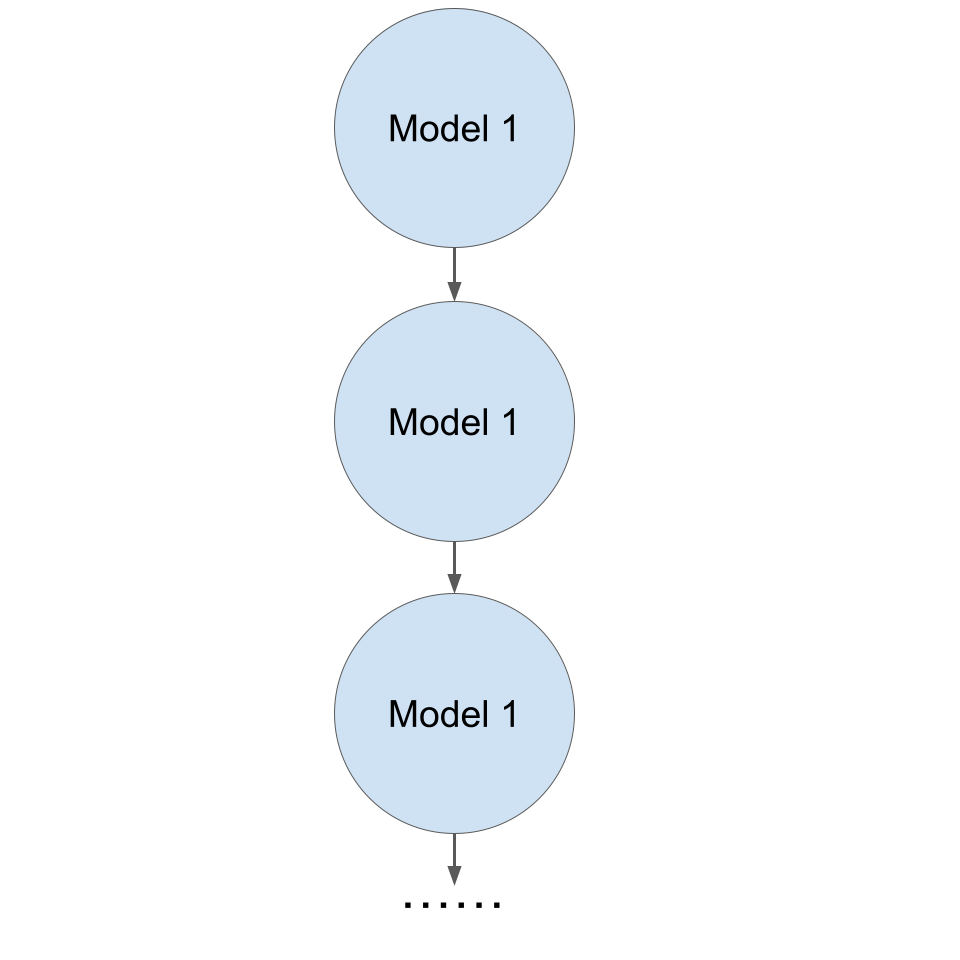}
  \caption{Self-labeling}
  \label{fig:self}
\end{subfigure}%
\begin{subfigure}{.5\textwidth}
  \centering
  \includegraphics[width=.8\linewidth]{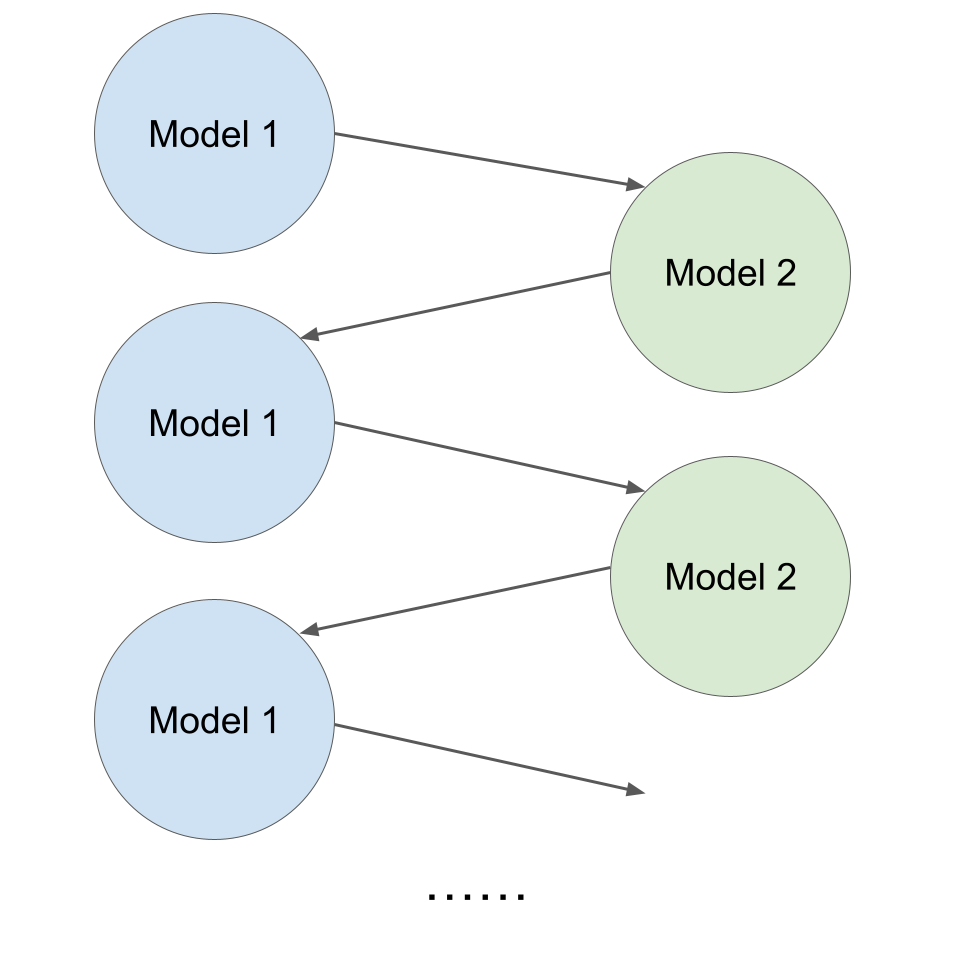}
  \caption{Cross-labeling}
  \label{fig:cross}
\end{subfigure}
 \bigskip
 \begin{subfigure}{.5\textwidth}
  \centering
  \includegraphics[width=.8\linewidth]{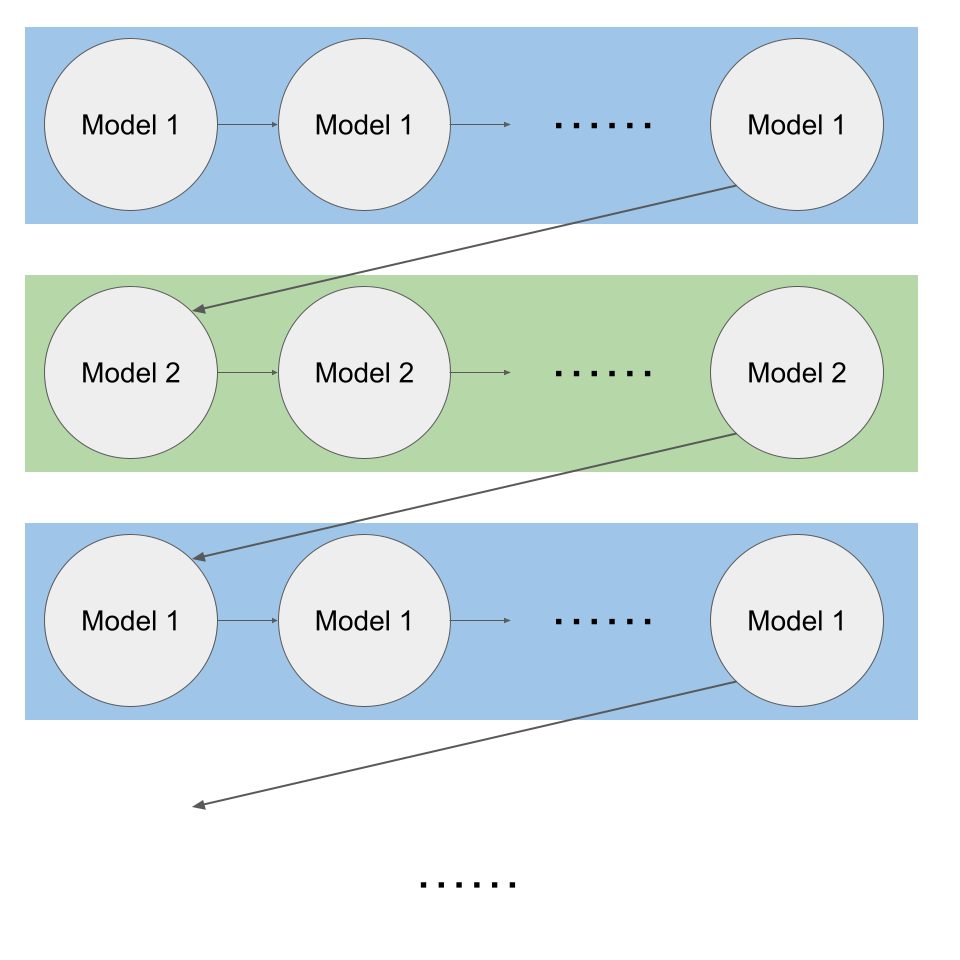}
  \caption{Joint cross- and self-labeling}
  \label{fig:cs}
\end{subfigure}%
\begin{subfigure}{.5\textwidth}
  \centering
  \includegraphics[width=.8\linewidth]{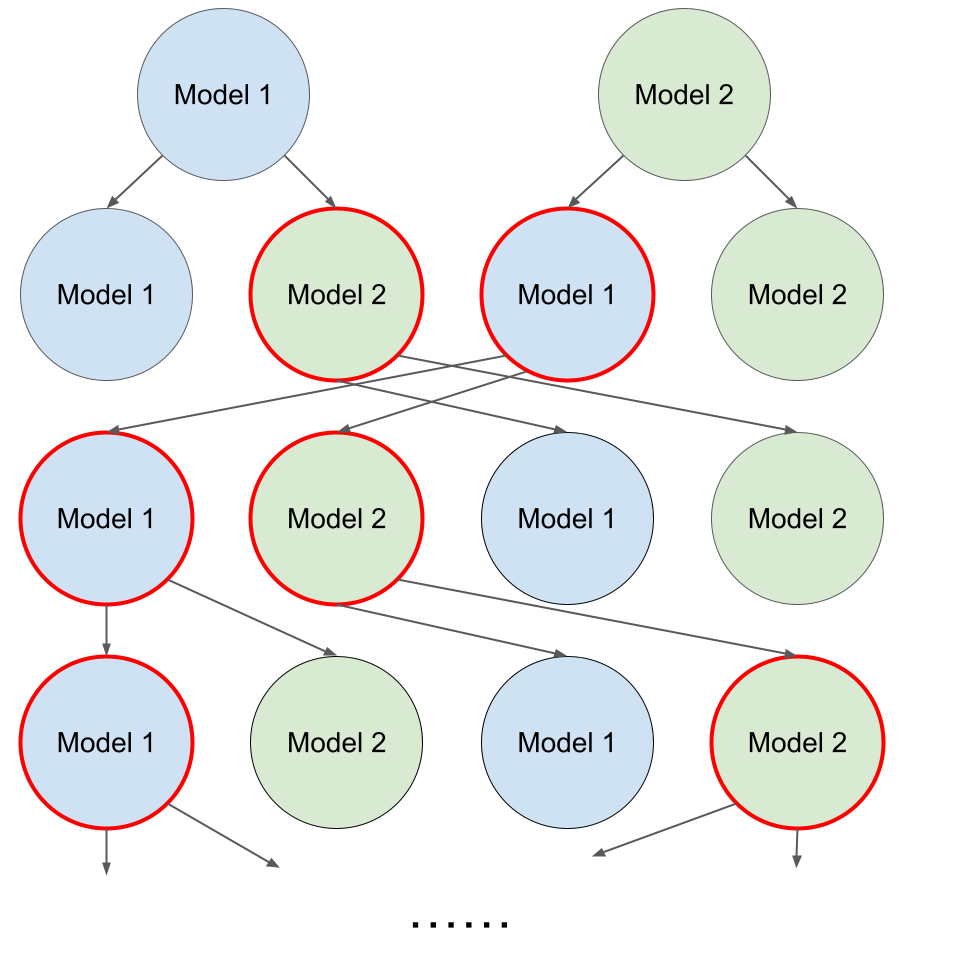}
  \caption{Greedy multi-labeling}
  \label{fig:greedy}
\end{subfigure}
\caption{Different GWS implementation for single-model and multi-model setting.}
\label{fig:model_demo}
\end{figure}

We provide a conceptual demonstration with two ranking model for all four implementations n Figure~\ref{fig:model_demo}. Note that the red circle in Figure~\ref{fig:greedy} means the best checkpoint in the iteration. To simplify the understanding, we only show the route starting from model 1 in Figure~\ref{fig:cross} and \ref{fig:cs}, but the route starting from model 2 is also conducted in parallel. 

The following subsections provide in-depth details and justification for all of these implementations of GWS. We first explain the initialization of these models, which is similar in all these four implementation. We then explain different re-labeling implementations. we also discuss the relationship between GWS and Expectation-Maximization. We show the notations used for the explanation in Table \ref{tab:not}.

\begin{table}[t]
\caption{Notations for GWS framework}
\label{tab:not}
\begin{tabular}{ll}
Notation & Definition \\ \hline
$Q$ & A query set  \\
$D$ & A document set \\
$\theta$ & Model parameters \\
$\theta^{(i)}$ & Model parameters of the $i$\textsuperscript{th} neural structure \\
$\theta^{(i,j)}$ & \makecell[l]{Model parameters of the $i$\textsuperscript{th} neural structure trained on\\ the triplets generated from the $j$\textsuperscript{th} neural structure}\\
$M$ & A ranking model \\
$M_{\theta}$ & A ranking model parameterized by $\theta$ \\
$T_{M}$ & A set of triplets generated by a model $M$ \\
$\theta^{(I)}$ / $M^{(i)}$ / $ T^{(i)}$ &  \makecell[l]{Model parameters / A model / A set of triplets from \\ the $i$\textsuperscript{th} neural structure.}
\end{tabular}
\end{table}

\subsection{Initialization in GWS}
The first step in GWS is to train the initial weakly supervised model using the typical weak supervision setup introduced by \citet{WS}. All the four re-labeling algorithms demonstrate that the initial weak labeling model is initialized by $\widehat{M}$, an existing unsupervised retrieval model, such as BM25 \cite{bm25}. 

Even though algorithms provide a general implementation of weak supervision, we only use the top $k$ retrieved documents by $\widehat{M}$ instead of all documents in the collection. This has been done for efficiency considerations. Thus, for every query $q_i \in Q$, let $\{d_{i1},d_{i2},...,d_{ik}\}$ be the top $k$ documents retrieved by $\widehat{M}$. Therefore, the training triplets for this query include $\{(q_i,d_{i1},\widehat{M}(q_i,d_{i1})),$ $\cdots,$ $(q_i,d_{ik},\widehat{M}(q_i,d_{ik}))\}$. Therefore, the initial training set $T_{\widehat{M}}$ consists of $|Q|\times k$ query-document pairs. Again, for efficiency reasons, we only re-score these documents in the following iterations of the GWS framework. In the following subsections, the training data used in iteration $t$ is denoted as $T_{M_{\theta_t}}$, which is generated from a model $M_{\theta_t}$ parameterized by $\theta_t$.  

\subsection{Iterative Re-Labeling and Training in GWS}
After building an initial model on weak supervision signals, we need to re-label the data by the current checkpoint, i.e., re-scoring all the triplets in the training data, and train a new model on these updated weak supervision signals.

In the $t$\textsuperscript{th} iteration, we optimize the model parameter $\theta_t$ based on the weakly supervised data $T_{M_{\theta_{t-1}}}$ generated from re-labeling in the previous iteration. Let the loss function be $\mathcal{L}(M,T)$ for the model $M$ and the data $T$, we have the following update in the training phase:
$$\theta_t = \arg \min_\theta \mathcal{L}(M_\theta, T_{M_{\theta_{t-1}}})$$
Note that the update is not related to $\theta_{t-1}$ since the operation is re-training a new model instead of fine-tuning the last parameter. We empirically found that starting from the initial model would lead to higher performance. The reason is that fine-tuning the last iteration is likely to overfit on the produced data. 

In the following, we introduce four re-labeling algorithms, described below. 

\subsubsection{Self-Labeling}
In Algorithm~\ref{alg:self-labeling}, we exploit the intermediate model as a new weak labeler to build new data for the next training iteration. Assume in the $t$\textsuperscript{th} training iteration, after training on weak supervision data $T_{t-1}$, we get a ranking model $M_{\theta_{t}}$. For the next iteration, we aim to build new weak supervision data $D_{t}$ by $M_{\theta_{t}}$. Regarding $M_{\theta_{t}}$ as the next weak labeler, we can update the relevance score in $D_{t-1}$. In summary, we have the following update in the self-labeling phase:
$$T_{M_{\theta_{t}}} =\{(q_i,d_{ij},M_{\theta_{t}}(q_i,d_{ij}))\} \text{ where }1\leq i \leq |Q|, 1\leq j \leq k$$
Again, we focus on the re-ranking problem in this work and only update scores for the same pairs as the previous data. 

Instead of re-labeling by only one model (i.e., self-labeling), we can use multiple weakly supervised models for re-labeling. The intuition is to increase the information diversity for the model in GWS. Because the training process runs on the same dataset by the same neural architecture, the over-fitting problem may deteriorate in iterative training. Thus, multi-model approaches aim to include different neural architectures in GWS to avoid this problem. In the following, we will introduce different implementations to let models be optimized and exchange their information. 

\begin{algorithm}[t]
\caption{Generalized Weak Supervision via Self-Labeling}
\label{alg:self-labeling}
\begin{algorithmic}[1]
\State \textbf{Input} (a) a set of queries $Q$; (b) a document collection $C$; (c) an unsupervised retrieval model $\widehat{M}$; (d) a loss function $\mathcal{L}$
\State \textbf{Output} a ranking model $M_\theta$. 
\State $M' \leftarrow \widehat{M}$
\Repeat
\State \textbf{Initialize} $\theta$.
\State $T \leftarrow \emptyset$ 
\For{$q \in Q$}
    \State $T \leftarrow T \cup M'(q, C)$
\EndFor
\State $\theta \leftarrow \arg \min_\theta \mathcal{L}(M_\theta, T)$ 
\State $M' \leftarrow M_\theta$
\Until{convergence}\\
\Return $M_\theta$
\end{algorithmic}
\end{algorithm}

\subsubsection{Cross-labeling}
As an alternative approach to self-labeling, Algorithm~\ref{alg:cross-labeling} aims to train $m$ ranking models at each iteration of GWS training and exchange the generated weak signals. In our experiments, we show that even the simplest case where $m=2$ improves self-labeling. That being said, Algorithm~\ref{alg:cross-labeling} can be used for any $m>1$ models. 

Without loss of generality, consider we have two ranking models parameterized by $\theta^{(1)}_t$ and $\theta^{(2)}_t$ at iteration $t$. In the re-labeling process of the $t$\textsuperscript{th} iteration, the two models generate two sets of weak supervision data, as follows:
$$T_{M_{{\theta_t^{(1)}}}} =\{(q_i,d_{ij},M_{{\theta_t^{(1)}}}(q_i,d_{ij}))\} \text{ where }1\leq i \leq |Q|, 1\leq j \leq k.$$
$$T_{M_{{\theta_t^{(2)}}}} =\{(q_i,d_{ij},M_{{\theta_t^{(2)}}}(q_i,d_{ij}))\} \text{ where }1\leq i \leq |Q|, 1\leq j \leq k.$$
During training, each model is trained on the data produced by the other model. Therefore, we have:
$$\theta_{t+1}^{(1)} = \arg\min_{\theta^{(1)}} \mathcal{L}(M_{\theta^{(1)}},T_{M_{{\theta_t^{(2)}}}})$$
$$\theta_{t+1}^{(2)} = \arg\min_{\theta^{(2)}} \mathcal{L}(M_{\theta^{(2)}},T_{M_{{\theta_t^{(1)}}}})$$
Through the operations, two models can exchange information during learning and avoid overfitting caused by self-labeling. In the case of two models, we easily set the supervision source as the other. For multiple models, we can choose a random one for each model and build one-to-one matching before training.


\begin{algorithm}[t]
\caption{Generalized Weak Supervision via Cross Labeling}
\label{alg:cross-labeling}
\begin{algorithmic}[1]
\State \textbf{Input} (a) a set of queries $Q$; (b) a document collection $C$; (c) an unsupervised retrieval model $\widehat{M}$; (d) a loss function $\mathcal{L}$.
\State \textbf{Output} $m$ ranking models $M_{\theta^{(1)}}, M_{\theta^{(2)}}, \cdots, M_{\theta^{(m)}}$. 
\State $M^{(1)}, M^{(2)}, \cdots, M^{(m)} \leftarrow \widehat{M}$
\Repeat
\State $T^{(1)}, T^{(2)}, \cdots, T^{(m)} \leftarrow \emptyset$ 
\State \textbf{Initialize} $\theta^{(1)}, \theta^{(2)}, \cdots, \theta^{(m)}$.
\For{$i \in [1, 2 \cdots, m]$}
    \For{$q \in Q$}
        \State $T^{(i)} \leftarrow T^{(i)} \cup M^{(i)}(q, C)$
    \EndFor
\EndFor
\For{$i \in [1, 2 \cdots, m]$}
    \State $\theta^{(i)} \leftarrow \arg \min_{{\theta}^{(i)}} \mathcal{L}(M_{\theta^{(i)}}, T^{(i-1)})$ 
    \State $M^{(i)} \leftarrow M_{\theta^{(i)}}$
\EndFor
\Until{convergence}\\
\Return $M_{\theta^{(1)}}, M_{\theta^{(2)}}, \cdots, M_{\theta^{(m)}}$
\end{algorithmic}
\end{algorithm}

\subsubsection{Joint Cross- and Self-labeling} 
Algorithm~\ref{alg:CLSL} combines self-labeling and cross-labeling settings. This approach still exchanges the generated weak signals among ranking models. However, it run a self-labeling process for each ranking model before exchanging labels.  Different from cross-labeling, Algorithm~\ref{alg:CLSL} aims to exchange the label from each model after convergence through self-labeling. As in the setting of cross-labeling, in the following, we only consider the simplest case where $m=2$, 


Without loss of generality, consider we have two ranking models parameterized by $\theta^{(1)}_t$ and $\theta^{(2)}_t$ at iteration $t$. In the re-labeling process at the $t$\textsuperscript{th} iteration, two models generate two sets of weak supervision data, as follows:
$$T_{M_{{\theta_t^{(1)}}}} =\{(q_i,d_{ij},M_{{\theta_t^{(1)}}}(q_i,d_{ij}))\} \text{ where }1\leq i \leq |Q|, 1\leq j \leq k.$$
$$T_{M_{{\theta_t^{(2)}}}} =\{(q_i,d_{ij},M_{{\theta_t^{(2)}}}(q_i,d_{ij}))\} \text{ where }1\leq i \leq |Q|, 1\leq j \leq k.$$
Following Algorithm~\ref{alg:self-labeling}, each model is trained on the data produced by itself. Therefore, we have:
$$\theta_{t+1}^{(1)} = \arg\min_{\theta^{(1)}} \mathcal{L}(M_{\theta^{(1)}},T_{M_{{\theta_t^{(1)}}}})$$
$$\theta_{t+1}^{(2)} = \arg\min_{\theta^{(2)}} \mathcal{L}(M_{\theta^{(2)}},T_{M_{{\theta_t^{(2)}}}})$$
Assume two models converge in the $L_1$\textsuperscript{th} and $L_2$\textsuperscript{th} iteration through the self-labeling process, we do an additional update to exchange the labels as the following:
$$\theta_{L_1+1}^{(1)} = \arg\min_{\theta^{(1)}} \mathcal{L}(M_{{\theta}^{(1)}},T_{M_{{\theta_{L_2}^{(2)}}}},)$$
$$\theta_{L_2+1}^{(2)} = \arg\min_{\theta^{(2)}} \mathcal{L}(M_{{\theta}^{(2)}},T_{M_{{\theta_{L_1}^{(1)}}}},)$$
After the exchange, we start a re-labeling process again as before.

\begin{algorithm}[t]
\caption{Generalized Weak Supervision via Joint Cross- and Self-Labeling}
\label{alg:CLSL}
\begin{algorithmic}[1]
\State \textbf{Input} (a) a set of queries $Q$; (b) a document collection $C$; (c) an unsupervised retrieval model $\widehat{M}$; (d) a loss function $\mathcal{L}$. 
\State \textbf{Output} $m$ ranking models $M_{\theta^{(1)}}, M_{\theta^{(2)}}, \cdots, M_{\theta^{(m)}}$. 
\State $M^{(1)}, M^{(2)}, \cdots, M^{(m)} \leftarrow \widehat{M}$
\Repeat
    \State $T^{(1)}, T^{(2)}, \cdots, T^{(m)} \leftarrow \emptyset$ 
    \For{$i \in [1, 2 \cdots, m]$}
    \For{$q \in Q$}
        \State $T^{(i)} \leftarrow T^{(i)} \cup M^{(i)}(q, C)$
    \EndFor
\EndFor
\For{$i \in [1, 2 \cdots, m]$}
    \State $\theta^{(i)} \leftarrow \arg \min_{{\theta}} \mathcal{L}(M_{\theta}, T^{(i-1)})$ 
    \State $M^{(i)} \leftarrow M_{\theta^{(i)}}$
\EndFor
    \For{$i \in [1, 2 \cdots, m]$}
        \State $M_{\theta^{(i)}}' \leftarrow \text{Algorithm 1}(Q, C, M_{(i)}', L)$
        \State $M_{(i)}' \leftarrow M_{\theta^{(i)}}$
    \EndFor
\Until{convergence}\\
\Return $M_{\theta^{(1)}}, M_{\theta^{(2)}}, \cdots, M_{\theta^{(m)}}$
\end{algorithmic}
\end{algorithm}

\begin{algorithm}[t]
\caption{Generalized Weak Supervision via Greedy Multi-Labeling}
\label{alg:greedy-cross}
\begin{algorithmic}[1]
\State \textbf{Input} (a) a set of queries $Q$; (b) a document collection $C$; (c) an unsupervised retrieval model $\widehat{M}$; (d) a loss function $\mathcal{L}$; (e) a validation error function V.
\State \textbf{Output} $m$ ranking models $M_{\theta^{(1)}}, M_{\theta^{(2)}}, \cdots, M_{\theta^{(m)}}$. 
\State $M^{(1)}, M^{(2)}, \cdots, M^{(m)} \leftarrow \widehat{M}$
\Repeat
\State $T^{(1)}, T^{(2)}, \cdots, T^{(m)} \leftarrow \emptyset$ 
\For{$i \in [1, 2 \cdots, m]$}
    \State \textbf{Initialize} $\theta^{(1,i)}, \theta^{(2,i)}, \cdots, \theta^{(m,i)}$.
\EndFor
\For{$i \in [1, 2 \cdots, m]$}
    \For{$q \in Q$}
        \State $T^{(i)} \leftarrow T^{(i)} \cup M^{(i)}(q, C)$
    \EndFor
\EndFor
\For{$i \in [1, 2 \cdots, m]$}
    \For{$j \in [1, 2 \cdots, m]$}
        \State $\theta^{(i,j)} \leftarrow \arg \min_{{\theta^{(i,j)}}} \mathcal{L}(M_{\theta^{(i,j)}}, T^{(j)})$ 
    \EndFor
    \State $G \leftarrow {\arg\min}_{j} V(\theta^{(i,j)})$
    \State $M^{(i)} \leftarrow  M_{\theta^{(i,G)}}$
\EndFor
\Until{convergence}\\
\Return $M_{\theta^{(1)}}, M_{\theta^{(2)}}, \cdots, M_{\theta^{(m)}}$
\end{algorithmic}
\end{algorithm}

\subsubsection{Greedy Multi-Labeling}
Greedy multi-labeling is a generalized version of cross-labeling. Different from choosing one fixed weak signal provider for each model as in Algorithms~\ref{alg:cross-labeling} and \ref{alg:CLSL}, we consider all possible $m$ models to build $m$ weak signal sets for one model structure, train $m$ checkpoints and pick the best one as the signal provider for the next iteration. In other words, at each iteration, we use all $m$ weak labelers as teachers and train all $m$ students and then select the best student models.

Consider we have two ranking models parameterized by $\theta^{(1)}_t$ and $\theta^{(2)}_t$ at iteration $t$. In the re-labeling of the $t$\textsuperscript{th} iteration, two models generate their own weak supervision data:
$$T_{M_{{\theta_t^{(1)}}}} =\{(q_i,d_{ij},M_{{\theta_t^{(1)}}}(q_i,d_{ij}))\} \text{ where }1\leq i \leq |Q|, 1\leq j \leq k.$$
$$T_{M_{{\theta_t^{(2)}}}} =\{(q_i,d_{ij},M_{{\theta_t^{(2)}}}(q_i,d_{ij}))\} \text{ where }1\leq i \leq |Q|, 1\leq j \leq k.$$
In the next iteration, each model need to be trained on all weak supervision data. Therefore, we have $m^2$ candidate models $\theta'$ as follows:
$$\theta^{(1,1)} = \arg \min_{{\theta^{(1)}}} \mathcal{L}(M_{\theta^{(1)}}, T_{M_{{\theta_t^{(1)}}}})$$
$$\theta^{(1,2)} = \arg \min_{{\theta^{(1)}}} \mathcal{L}(M_{\theta^{(1)}}, T_{M_{{\theta_t^{(2)}}}})$$
$$\theta^{(2,1)} = \arg \min_{{\theta^{(2)}}} \mathcal{L}(M_{\theta^{(2)}}, T_{M_{{\theta_t^{(1)}}}})$$
$$\theta^{(2,2)} = \arg \min_{{\theta^{(2)}}} \mathcal{L}(M_{\theta^{(2)}}, T_{M_{{\theta_t^{(2)}}}})$$
For each model structure, we choose the best one as the weak signal provider in the next iteration as the following:
$$\theta_{t+1}^{(1)} = {\arg\min}_{\theta \in \{\theta^{(1,1)}, \theta^{(1,2)}\}} V(\theta)$$
$$\theta_{t+1}^{(2)} = {\arg\min}_{\theta \in \{\theta^{(2,1)}, \theta^{(2,2)}\}} V(\theta)$$
Where $V$ is the validation error function for the candidate models. Note that we still need a small validation set to judge which checkpoints are the best ones for our tasks. We leave the fully-unsupervised judgment to the future extension.




\subsection{Relationship of GWS and Expectation-Maximization}
To better understand the theoretical foundation of GWS, we draw connection between GWS and Expectation-Maximization (EM) which has been widely explored in various machine learning tasks. To this aim, we need to revisit GWS from the probabilistic view. For simplicity, this section focuses on the self-labeling approach (Algorithm \ref{alg:self-labeling}). Let $R \in \{0, 1\}$ be a binary random variable that represents whether a document is relevant to a query or not. Thus, self-labeling is equivalent to inferring labels based on $P(R=1|q,d;\theta)$ and $P(R=0|q,d;\theta)$. For the iterative training, minimizing a loss function $\mathcal{L}(M_\theta, T)$ can be considered as the negative log-likelihood for the current relevance judgment in $T$.

Now let us focus on the EM algorithm, a general learning framework for unsupervised learning problems. Given a joint distribution $P(X,Z|\theta)$, where $X$ is the observed data, $Z$ is the hidden or missing variable and $\theta$ is a set of model parameters, the EM algorithm aims to maximize $P(X|\theta)$ by the following steps:
\begin{enumerate}
    \item Initialize $\theta_0$
    \item E-step: Estimate $Z$ by $P(Z|X,\theta_{t-1})$
    \item M-step: $\theta_t = {\arg\min}_{\theta}-P(Z|X,\theta_{t-1})\log P(X,Z|\theta)$
    \item Repeat step 2 and 3 until it converges.
\end{enumerate}

Comparing the E-step and M-step of the EM algorithm with the probabilistic view of self-labeling and iterative training, it is clear that the process of GWS could be connected to EM if we regard $R$ as the hidden variable $Z$; and $Q$ and $D$ as the observed data $X$. In other words, in Algorithm \ref{alg:self-labeling}, lines 6-9 can be connected to the E-Step and Lines 10-11 can be connected to the M-Step of the EM algorithm. However, the GWS framework behaves differently for initialization, which significantly affects the performance of the model.

The result of EM algorithm is always affected by the initialization of parameters. For retrieval tasks, random initialization on hidden variables (as often done in the EM algorithms), which are relevance judgments, is not applicable because relevance judgment is always complex and imbalanced in the collection. On the other hand, training a ranking model on randomly generated ranked lists may not converge to an effective parameter setting for ranking tasks.

We regard our process as an EM process with a weak supervision initialization. Through weak supervision signals, we get non-random initialization and have noisy but useful results for the first expectation step. Therefore, the following EM process has an excellent base to generate a feasible model for ranking tasks.

For the other three labelings with multi-model settings, they could not be directly linked to EM process, but we consider them as a more generic process than EM. Our experiments also show that they perform better than self-labeling which is equivalent to EM.

\subsection{Loss Function in GWS}
Following the empirical results presented by \citet{WS} and the theoretical results presented by \citet{theoryWS}, we use a pairwise loss function for optimizing GWS models. However, as is also shown in the experiments, we observed that existing loss functions are not sufficiently effective for GWS optimization.
Because the initial weak labeler is imperfect, the poor performance of the weak labeler on some queries is inevitable. If we assume all queries have the same importance through our training process, the poor performing queries are expected to have negative impact on the final performance. To keep up the quality of initial weak supervision data, we assign a weight to each query based on its estimated ranking performance and integrate it into our optimization.

Assume for each query $q$, the corresponding importance is $w_q$.  We can use in-batch re-weighting to normalize the importance for each training instance. For each training batch $B = \{(q_1,d_{q_1,1},d_{q_1,2}),\\ (q_2,d_{q_2,1},d_{q_2,2}), \cdots, q_{|B|},d_{q_{|B|},1},d_{q_{|B|},2})\}$, our loss function is defined as follows:
\begin{align*}
    l(B) &= \sum_{i=1}^{|B|}{l(q_i,d_{q_i,1},d_{q_i,2};M_\theta, M')} \\
    &= \sum_{i=1}^{|B|}{\frac{w_{q_i}}{\sum_{j=1}^{|B|}{w_{q_j}}} l_{\text{hinge}}(q_i,d_{q_i,1},d_{q_i,2};M_\theta, M')} \\
    &= \sum_{i=1}^{|B|}{\frac{w_{q_i}}{\sum_{j=1}^{|B|}{w_{q_j}}} \max\left(0,\epsilon-\text{sign}(M'(q_i,d_{q_i,1})-M'(q_i,d_{q_i,2})) (M_\theta(q_i,d_{q_i,1})-M_\theta(q_i,d_{q_i,2}))\right)}
\end{align*}
where $l_{\text{hinge}}(q_i,d_{q_i,1},d_{q_i,2};M_\theta, M')$ is the hinge loss for the pairwise training instance $(q_i,d_{q_i,1},d_{q_i,2})$ and the ranking model $M_\theta$. The labels come from the weak labeler $M'$. In hinge loss, $\epsilon$ is a margin. We set $\epsilon=1$ in our experiment.



For estimating query weights, we rely on query performance prediction (QPP). The goal of QPP is to predict a retrieval model's quality for a given query when neither explicit nor implicit relevance information is available \cite{clarity}. Thus, we can leverage unsupervised QPP models for estimating the quality of a ranked list produced by the weak labeler during training and filter out noisy data in the weak supervision signal. 

Among all the available QPP methods, we choose Normalized Query Commitment (NQC) \cite{NQC} as our QPP estimator, because its robust performance and its simplicity. That being said, the choice of QPP method is orthogonal to the GWS optimization process and it can be replaced by any other QPP method. NQC estimates the retrieval performance by computing the normalized standard deviation of the retrieval scores assigned to the top retrieved documents. The formula is as follows:
$$NQC(q;C,M') = \frac{\sqrt{\frac{1}{n}\sum_{d\in \pi^{k}_{M'}(q;C)}(\text{score}(q,d)-\mu)^2}}{\text{score}(q,C)},$$
where $\pi^{k}_{M'}(q;C)$ is the top $k$ documents retrieved by the retrieval model $M'$ (which is the weak labeler in our case) in response to query $q$. $\mu$ is the average of the scores in $\pi^{k}_{M'}(q;C)$. $\text{score}(q,C)$ concatenates all documents in the collection and computes the relevance score. In this work, we directly adopt $NQC(q;C,M')$ to estimate $w_q$. For the ranking models based on pre-train language models, we cannot compute $\text{score}(q,C)$, so we ignore this normalization term for them in the experiment, and it does not affect our computation for $l(B)$. 

For optimization, we adopt the batch stochastic gradient descent algorithm. For each batch, we compute the average loss over all document pairs in the batch and update the parameters.

\section{Experiment}
In this section, we introduce the experiments and discuss the results. We give a description of two datasets we used, explain the evaluation metrics, show the detail of our experimental setup and discuss the results and additional analysis.

\label{table:data}
\begin{table}[t]
    \centering
    \caption{Statistic of WikiPassageQA and ANTIQUE datasets.} 
    \label{tab:anti}
    \begin{tabular}{lll}\Xhline{1.5pt}
&\textbf{ANTIQUE}& \textbf{WikiPassageQA} \\\Xhline{1.5pt}
\# training queries & 2,466  & 3332        \\
\# validation queries & -  & 417        \\
\# testing queries & 200 & 416         \\
\# training docs  & 27,422 & 194314        \\
\# validation docs & 2,466  & 25841        \\
\# testing docs  & 6589 & 23981         \\
\# terms/query           & 10.51     & 9.52       \\
\# terms/document             & 47.75 & 133.092  \\\midrule
\# label 3 & 13067 &    -     \\
\# label 2  & 9276  &  -      \\
\# label 1           & 8754 &  6260             \\
\# label 0             & 2914 & 212035 \\\Xhline{1.5pt}
\end{tabular}

\end{table}

\subsection{Data}
In our experiments, we use two datasets for evaluation. The first one is \textbf{ANTIQUE}, which is a passage retrieval dataset for non-factoid questions, created by \citet{antique} based on Yahoo! Webscope L6. Relevance annotations are collected through crowdsourcing based on the standard pooling technique. Relevance labels are between 0 and 3. The statistics of this dataset are presented in Table \ref{tab:anti}.
The second dataset is \textbf{WikiPassageQA} \cite{wikipassqa}, which is a passage retrieval dataset from Wikipedia articles for questions generated through crowdsourcing. WikiPassageQA provides binary relevance labels. The statistics of this dataset are also reported in Table \ref{tab:anti}. 

Note that, given the focus of this paper on weak supervision, none of the relevance judgments are used for training. 


\subsection{Evaluation Metrics}
To evaluate retrieval effectiveness, we report four standard evaluation metrics: (1, 2) normalized discounted cumulative gain (NDCG) at two ranking cut-offs 1 and 10. NDCG is a standard metric that considers graded relevance labels. (3) Mean reciprocal rank (MRR) that measures the reciprocal rank of the first relevant retrieved document, and (4) mean average precision (MAP) that is a standard recall-oriented metric introduced by TREC. We only consider the documents in the re-ranking scope for measuring MAP; 
As mentioned earlier, ANTIQUE provides four-level graded relevance annotation, while the last two metrics (MRR and MAP) only take binary labels. To convert graded relevance labels to binary labels, we followed the instructions provided by the ANTIQUE dataset: labels 0 and 1 are non-relevant and labels 2 and 3 are relevant. 

Statistically significant differences in metric values are determined using the two-tailed paired t-test with Bonferroni correction and 95\% confidence interval ($p\_value < 0.05$).

\begin{table}[t]
\caption{The retrieval performance obtained by GWS and the baselines. The superscripts * respectively denote that the improvements over the weakly supervised models are statistically significant. The highest value in each column of the table is marked in bold.}
\label{table:main}
\begin{tabular}{lcccc}
\Xhline{1.5pt}
\multicolumn{5}{c}{\textbf{WikiPassageQA}} \\ \Xhline{1.5pt}
\textbf{Model} & \textbf{NDCG@1} & \textbf{NDCG@10} & \textbf{MAP} & \textbf{MRR} \\ \hline
\textbf{\texttt{Baselines}} \\ 
\quad BM25 (initial weak labeler) & 0.4087 & 0.5374 & 0.4685 & 0.5479 \\
\quad BERT - Zero Shot & 0.0337 & 0.1102 & 0.1115 & 0.1146 \\
\quad RoBERTa - Zero Shot & 0.0601 & 0.1485 & 0.1373 & 0.1534 \\
\quad BERT - WS & 0.4928 & 0.6345 & 0.5574 & 0.6379 \\
\quad RoBERTa - WS & 0.5000 & 0.6316 & 0.5879 & 0.6692 \\ \hline
\textbf{\texttt{GWS with Self-Labeling}} \\ 
\quad BERT - Self & 0.5553* & 0.6895* & 0.6116* & 0.6938* \\
\quad RoBERTa - Self & 0.6058* & 0.7310* & 0.6588* & 0.7413* \\\hline
\textbf{\texttt{GWS with Cross-Labeling}} \\ 
\quad BERT - Cross & 0.5745* & 0.7052* & 0.6307* & 0.7097* \\
\quad RoBERTa - Cross & 0.5673* & 0.7007* & 0.6248* & 0.7042* \\ \hline
\textbf{\texttt{GWS with Joint Cross- and Self-Labeling}} \\ 
\quad BERT - JCS & \textbf{0.6611}* & 0.7669* & \textbf{0.6995}* & \textbf{0.7774}* \\
\quad RoBERTa - JCS & {0.6394}* & 0.7519* & 0.6836* & 0.7653* \\ \hline
\textbf{\texttt{GWS with Greedy Multi-Labeling}} \\ 
\quad BERT - Multi & {0.6490}* & 0.7492* & 0.6805* & 0.7683* \\
\quad RoBERTa - Multi & {0.6394}* & \textbf{0.7685}* & 0.6787* & 0.7630* \\ \Xhline{1.5pt}

\multicolumn{5}{c}{\textbf{ANTIQUE}} \\ \Xhline{1.5pt}
\textbf{\texttt{Baselines}} \\ 
\quad BM25 (initial weak labeler) & 0.4417 & 0.3675 & 0.1540 & 0.5277 \\
\quad BERT - Zero Shot & 0.3867 & 0.3591 & 0.1494 & 0.4818 \\
\quad RoBERTa - Zero Shot & 0.2783 & 0.2727 & 0.1123 & 0.3797 \\
\quad BERT - WS & 0.4967 & 0.3981 & 0.1753 & 0.5794 \\
\quad RoBERTa - WS & 0.4617 & 0.3776 & 0.1652 & 0.5706 \\ \hline
\textbf{\texttt{GWS with Self-Labeling}} \\ 
\quad BERT - Self & 0.5383* & 0.4202* & 0.1863* & 0.6300* \\
\quad RoBERTa - Self & 0.5917* & 0.4270* & 0.1923* & 0.6648* \\\hline
\textbf{\texttt{GWS with Cross-Labeling}} \\ 
\quad BERT - Cross & 0.5717* & 0.4285* & 0.1930* & 0.6446* \\
\quad RoBERTa - Cross & 0.5833* & 0.4246* & 0.1941* & 0.6645* \\\hline
\textbf{\texttt{GWS with Joint Cross- and Self-Labeling}} \\ 
\quad BERT - JCS & 0.5833* & 0.4303* & 0.1887* & 0.6488* \\
\quad RoBERTa - JCS & 0.6067* & 0.4270* & 0.1936* & 0.6745* \\\hline
\textbf{\texttt{GWS with Greedy Multi-Labeling}} \\ 
\quad BERT - Multi & 0.5867* & \textbf{0.4337}* & 0.1942* & 0.6509* \\
\quad RoBERTa - Multi & \textbf{0.6250}* & 0.4327* & \textbf{0.1957}* & \textbf{0.6851}*
\\ \Xhline{1.5pt}
\end{tabular}
\end{table}

\subsection{Experimental Setup}
GWS is a framework which is compatible with any ranking architectures and initial weak labelers. In this part, we describe the actual our experimental setup of GWS. Following \citet{WS}, we choose BM25 as the initial weak labeler, which has shown robust and strong performance across collections. In our experiments, we use the Anserini's implementation of BM25 \cite{anserini}. For the ranking architecture, we choose two pre-trained language models, BERT \cite{BERT} and RoBERTa \cite{roberta}. Recently, fine-tuning BERT for ranking tasks has received notable attention \cite{bertrank1,bertrank2}. Compared to a neural ranking model trained from scratch, BERT and other language models improve the ranking performance significantly. Besides, fine-tuning a pre-trained language model also decreases the required volume of weak supervision data.

For the input of BERT, we concatenate a query and a document with a \texttt{[SEP]} token to compute their relevance. For the text-matching task, the pooled output of BERT (the encoding of [CLS] token) would be fed into a feed-forward network to compute a matching score. For the score, we can compute the loss function for ranking and fine-tune the parameters according to the loss. Fine-tuning BERT and optimizing the final feed-forward network with the ranking loss function is a general method to apply BERT for learning to rank. RoBERTa has the same usage as BERT.

All ranking models are implemented by PyTorch \cite{pytorch} and the HuggingFace Transformer library \cite{transformers}. For the pre-trained language models used in our experiments, we used the checkpoints for BERT-base \cite{BERT} and RoBERTa-base \cite{roberta} implementations of HuggingFace. 
For optimization, we adopt the AdamW optimizer \cite{adamW} with the initial learning rate of $5 \times 10^{-5}$, $\beta_1=0.9$, $\beta_2=0.99$, and weight decay of $0.01$. we set the batch size as $16$, and the total training steps as $10000$.

For teacher/student model selection, we rely on the performance on a held-out validation set. For WikiPassageQA, we use the original development set for validation. However, ANTIQUE does not have an explicit validation set. Thus, we randomly select 10\% of training queries as the validation set. We check the performance on the validation set every 1000 steps and use the best one as the final model. The validation sets are used for all our models and all the baseline methods. 

For re-ranking tasks, we need to decide the number of documents to be considered for re-ranking. WikiPassageQA provides an explicit set of documents to be re-ranked for each query. For ANTIQUE, given the nature of the dataset, we re-rank the top 20 documents retrieved by BM25. The same setting is used for all methods, including baselines. To create weakly supervised dataset for training, we created 20 pairs of documents per query. Unlike random sampling on arbitrary pairs, we adopt a policy that regards only the top-half passages in the list as positive and the other half as negative samples. We randomly pick one passage from both sets to build a training pair.

\subsection{Results and Discussion}

In this section, we report and discuss the results obtained from GWS models and the baselines on two passage retrieval datasets.


\paragraph{\textbf{Baseline Results.}} We compare GWS with three sets of baselines: (1) the initial weak labeler (i.e., Anserini's BM25), (2) the BERT and RoBERTa ranking models under the zero-shot setting, and (3) the BERT and RoBERTa models fine-tuned using the original weak supervision approach of \citet{WS}.  Table \ref{table:main} presents the results for the baselines and the models trained via GWS. 

As was also discovered by other researchers \cite{zeroshot}, large language models, such as BERT and RoBERTa, have a poor zero-shot retrieval performance; thus, fine-tuning them with a retrieval objective is necessary. Zero-shot retrieval has meaningful results on ANTIQUE because we focus on only top-20 re-ranking from BM25 results, and most of them have relevance scores from 1-3, contributing to evaluation metrics.

That being said, once these models are fine-tuned using weakly supervised data (i.e., BERT - WS and RoBERTa - WS), they substantially outperform their weak labeler (i.e., BM25) without any manually labeled data. The observed improvements compared to BM25 are larger in WikiPassageQA. For example, BERT - WS outperforms BM25 by 18\% and by 8\% in terms of NDCG@10 on WikiPassageQA and ANTIQUE datasets, respectively. This once again confirms the power of weak supervision training for neural ranking models that was originally discovered by \citet{WS}.   In the weak supervision setting, there is no clear winner between BERT and RoBERTa models; RoBERTa performs better on WikiPassageQA, especially in terms of MAP and MRR, while BERT outperforms RoBERTa on ANTIQUE with respect to all metrics. 

\paragraph{\textbf{GWS with Self-Labeling Results.}}
Results on both datasets confirm that GWS with Self-Lebeling outperforms all the baselines, including weakly supervised models. For example, a BERT model that is initially trained on the BM25's weak labels and then uses the proposed self-labeling and iterative training strategy achieves 8\% and 5.5\% higher NDCG@10 values than BERT - WS. These improvements, for both BERT and RoBERTa models, are statistically significant. Therefore, \textit{we can conclude that GWS with Self-Labeling leads to retrieval performance improvements in all cases}. It is notable that most impacted evaluation metrics by self-labeling are NDCG@1 and MRR. This suggests that self-labeling most impact the model's behavior in identifying the first relevant document at top positions. These metrics are often important in non-factoid question answering tasks.
Interestingly, RoBERTa benefits more from self-labeling; RoBERTa - Self outperforms RoBERTa - WS by 16\% and 13\% in terms of NDCG@10 on WikiPassageQA and ANTIQUE datasets, respectively. Obtaining such substantial improvements without using labeled training data is the first evidence for potential impacts of GWS.

\begin{figure}[h]
\begin{subfigure}{.35\textwidth}
  \centering
    \begin{subfigure}{\textwidth}
	   \includegraphics[width=\textwidth]{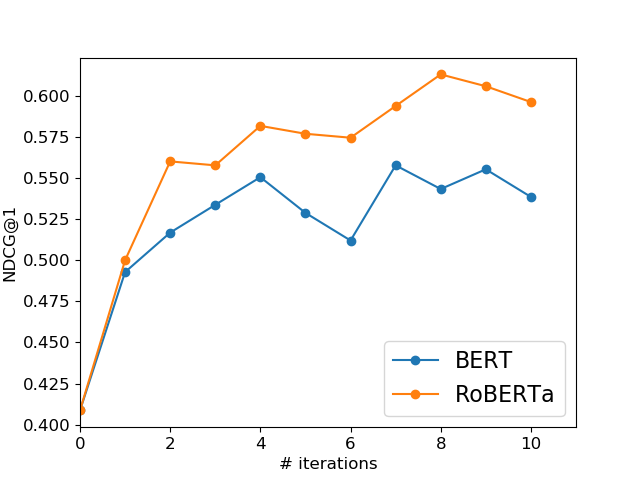}
	   \caption{NDCG@1}
    \end{subfigure}
    \begin{subfigure}{\textwidth}
	   \includegraphics[width=\textwidth]{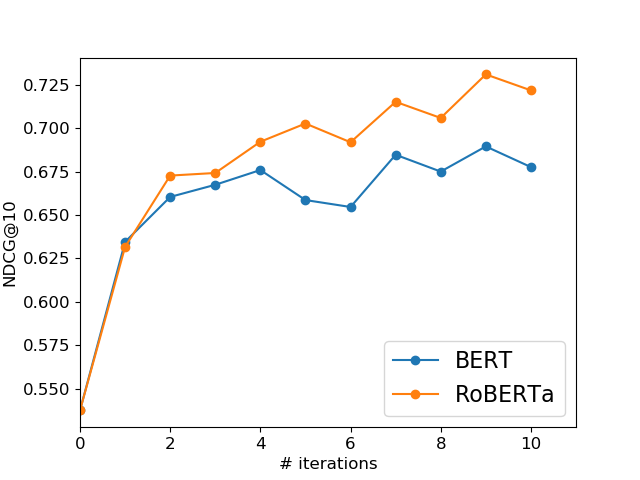}
	   \caption{NDCG@10}
    \end{subfigure}
    \begin{subfigure}{\textwidth}
	   \includegraphics[width=\textwidth]{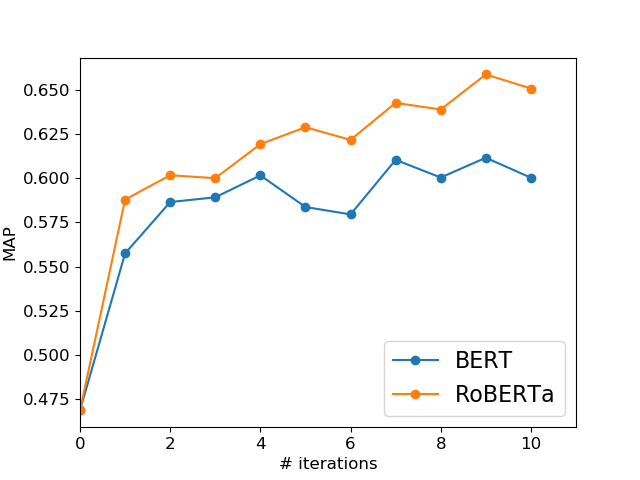}
	   \caption{MAP}
    \end{subfigure}
    \begin{subfigure}{\textwidth}
	   \includegraphics[width=\textwidth]{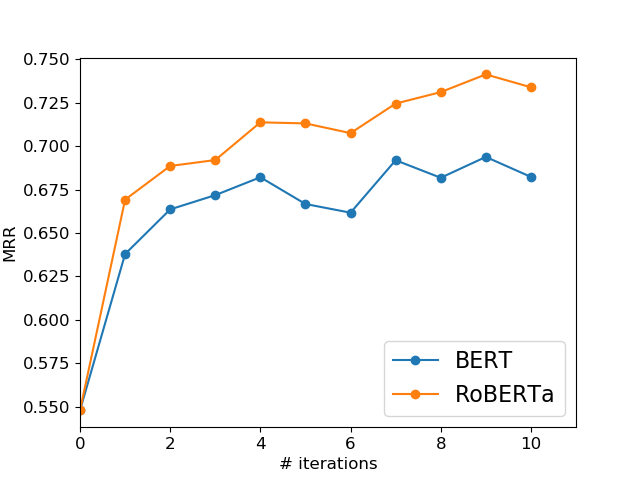}
	   \caption{MRR}
    \end{subfigure}
\end{subfigure}
\begin{subfigure}{.35\textwidth}
  \centering
    \begin{subfigure}{\textwidth}
	   \includegraphics[width=\textwidth]{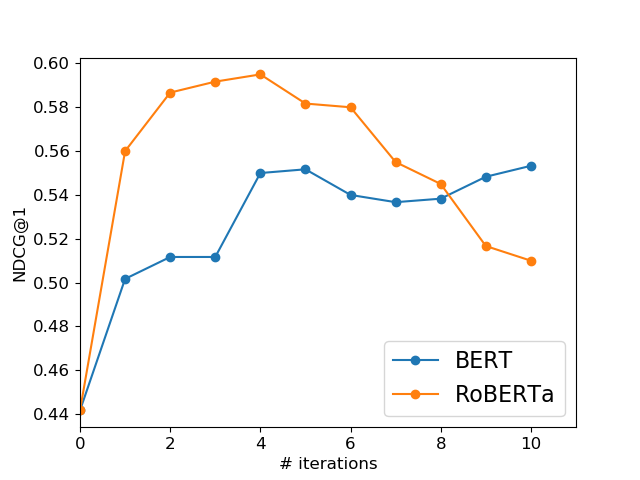}
	   \caption{NDCG@1}
    \end{subfigure}
    \begin{subfigure}{\textwidth}
	   \includegraphics[width=\textwidth]{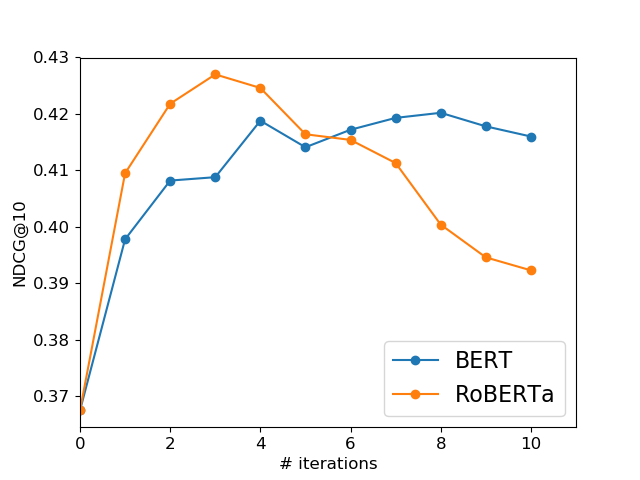}
	   \caption{NDCG@10}
    \end{subfigure}
    \begin{subfigure}{\textwidth}
	   \includegraphics[width=\textwidth]{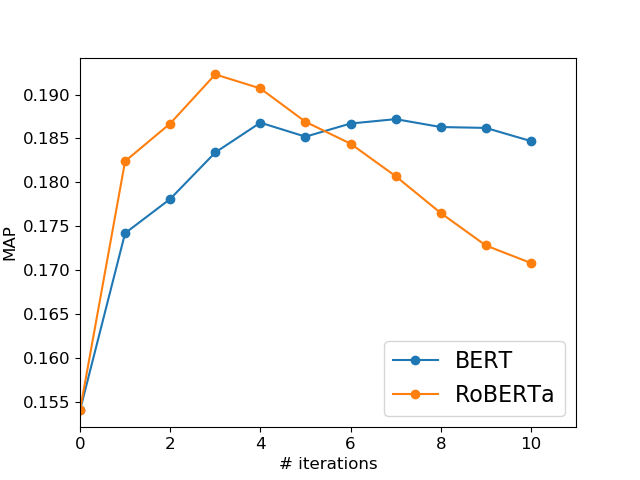}
	   \caption{MAP}
    \end{subfigure}
    \begin{subfigure}{\textwidth}
	   \includegraphics[width=\textwidth]{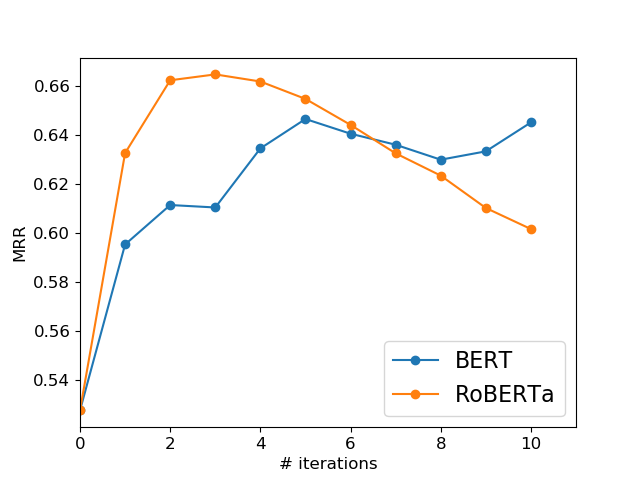}
	   \caption{MRR}
    \end{subfigure}
\end{subfigure}
\caption{The retrieval performance obtained by GWS with self-labeling at different iterations. Results are presented on both WikiPassageQA ((a)-(d)) and ANTIQUE ((e)-(h)) datasets. Iteration 0 denotes the weak labeler's performance.}
\label{fig:curve}
\end{figure}

To better understand the behavior of GWS with self-labeling, we plot a curve of ranking performance at each re-labeling iteration. The results are depicted in Figure \ref{fig:curve}. Note that the results for iteration 0 come from the initial weak labeler, BM25 in our experiment. The results for iteration 1 are equivalent to results obtained by the original weak supervision approach.
In WikiPassageQA, we observe that the ranking performance generally increases through the iterations. Although the curve sometimes drops, both models reach the best performance after iteration 6 on all metrics, which highlights the importance of iterative re-labeling. Overall, the both BERT and RoBERTa curves follow a similar trend on WikiPassageQA.  Results in ANTIQUE are different. Both BERT and RoBERTa reach their best performance in the early iterations. The performance curves for BERT remains stable in the following iterations, however, RoBERTa observes a substantial performance drop in late iterations. That being said, RoBERTa at its best performing iteration outperforms BERT. Besides the importance of self-labeling, these plots suggest that the iterative optimization behavior in GWS is dataset-dependent and sometimes an early stopping approach is needed. Therefore, a validation set for determining the best performing iteration may play a vital role.





\paragraph{\textbf{GWS with Cross-Labeling Results.}}
From Table~\ref{table:main}, we observe that GWS with cross-labeling significantly outperforms all the baselines. Compared to self-labeling, cross-labeling does not provide a consistant improvement. For example, BERT with cross-labeling outperforms BERT with self-labeling on both datasets, however, this is not the case for RoBERTa. RoBERTa learns better from self-labeling for WikiPassageQA and both self-labeling and cross-labeling strategies have a comparable impact on RoBERTa for the ANTIQUE datasets. One reason may be that RoBERTa plays a better role as a teacher model, thus whenever it's a teacher, either as a RoBERTa - Self or BERT - Cross, it leads to superior performance. 


\paragraph{\textbf{GWS with Joint Cross- and Self-Labeling Results.}}
Joint Cross- and Self-Lebeling (JCS) demonstrates a successful performance compared to the previous implementations of GWS. The reason is that JCS combines the benefits of both self-labeling and cross-labeling; self-labeling brings us performance robustness by iterative re-labeling and optimization, while cross-labeling better enables knowledge transfer and prevents over-fitting. The improvements brought by JCS are higher in WikiPassageQA. There is no clear winner between BERT - JCS and RoBERTa - JCS; BERT - JCS performs better on WikiPassageQA, while RoBERTa - JCS does well on the ANTIQUE dataset. 

\paragraph{\textbf{GWS with Greedy Multi-Labeling Results.}}
The results obtained by the Greedy Multi-Labeling approach is consistent with JCS. This approach performs better than each of the self-labeling and cross-labeling approaches, separately. This approach takes the best student model at every iteration as the teacher for the next iteration, therefore, similar to JCS, this approach also combines the benefits of both self-labeling and cross-labeling. In fact, the best performing GWS approaches are JCS and Greedy Multi-Labeling. On WikiPassageQA, our best performing model outperforms the initial weak labeler (BM25) by 56\% and 43\% in terms of NDCG@1 and NDCG@10, respectively. On ANTIQUE, the improvements are slightly smaller; our best performing model respectively outperforms the initial weak labeler by 41\% and 18\% in terms of NDCG@1 and NDCG@10.



\begin{figure}[t]
\begin{subfigure}{.45\textwidth}
  \centering
  \label{subfig:NDCG1_ana}
	\includegraphics[width=\textwidth]{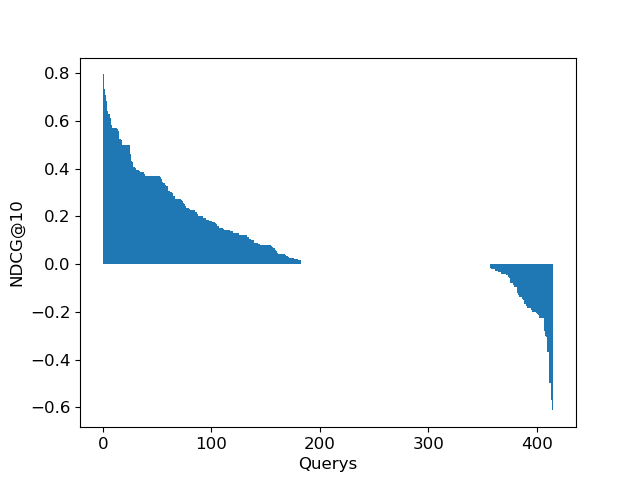}
	\caption{NDCG@10 on WikiPassageQA}
\end{subfigure}%
\begin{subfigure}{.45\textwidth}
  \centering
  \label{subfig:NDCG10_ana}
	\includegraphics[width=\textwidth]{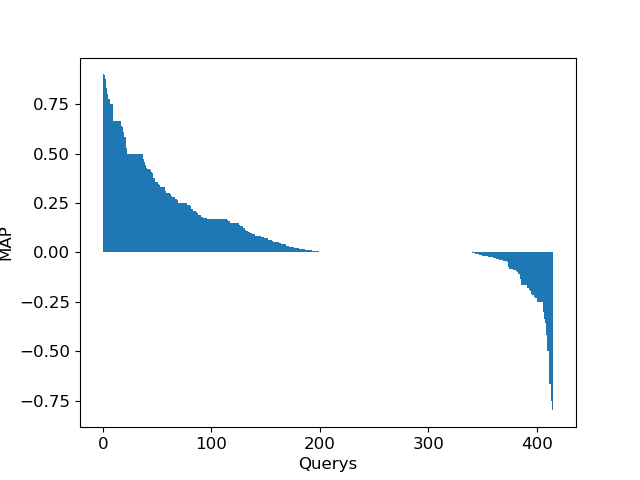}
	\caption{MAP on WikiPassageQA}
\end{subfigure}%

\begin{subfigure}{.45\textwidth}
  \centering
  \label{subfig:MAP_ana}
	\includegraphics[width=\textwidth]{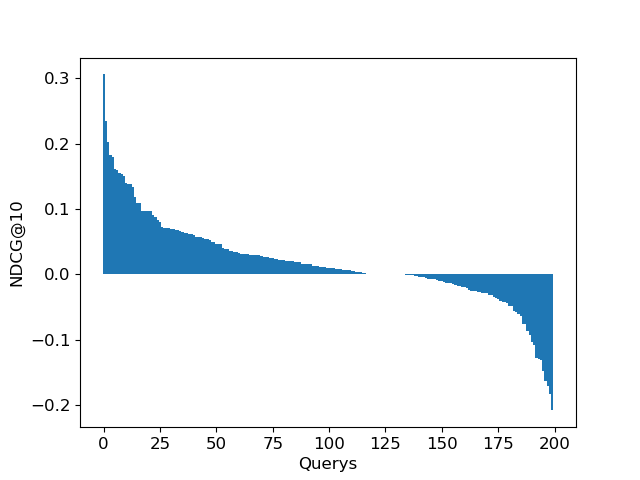}
	\caption{NDCG@10 on ANTIQUE}
\end{subfigure}%
\begin{subfigure}{.45\textwidth}
  \centering
  \label{subfig:MRR_ana}
	\includegraphics[width=\textwidth]{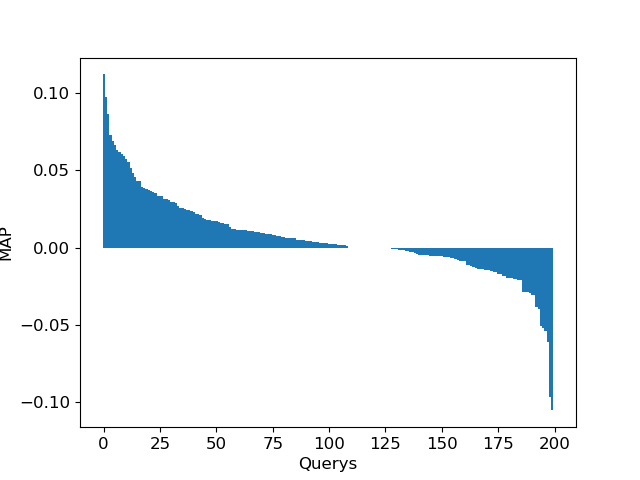}
	\caption{MAP on ANTIQUE}
\end{subfigure}%
\caption{The difference of ranking performance between RoBERTa - GWS (self) and RoBERTa - WS over all queries in terms of NDCG@10 and MAP on WikiPassageQA((a)-(b)) and ANTIQUE((c)-(d)) .}
\label{fig:diff}
\end{figure}

\begin{table}[t]
\caption{The impact of query importance weighting in GWS training on the retrieval performance. The superscript * denotes that the improvements obtained by query importance weighting are statistically significant.}
\label{table:ablation}
\begin{tabular}{lcccc}
\Xhline{1.5pt}
\multicolumn{5}{c}{\textbf{WikiPassageQA}} \\ \Xhline{1.5pt}
\textbf{Model} & \textbf{NDCG@1} & \textbf{NDCG@10} & \textbf{MAP} & \textbf{MRR} \\ \hline
 BERT - Self w/o NQC & 0.5337 & 0.6637 & 0.5910 & 0.6785 \\
 BERT - Self & 0.5553 & 0.6895 & 0.6116 & 0.6938 \\
 RoBERTa - Self w/o NQC & 0.5889 & 0.6889 & 0.6183 & 0.7099 \\ 
 RoBERTa - Self & 0.6058 & 0.7310 & 0.6588 & 0.7413 \\\Xhline{1.5pt}

\multicolumn{5}{c}{\textbf{ANTIQUE}} \\ \Xhline{1.5pt}
 BERT - Self w/o NQC & 0.5300 & 0.4066 & 0.1863 & 0.6188 \\
 BERT - Self & 0.5383 & 0.4202 & 0.1863 & 0.6300 \\
 RoBERTa - Self w/o NQC & 0.5433 & 0.4059 & 0.1867 & 0.6306 \\ 
 RoBERTa - Self & 0.5917 & 0.4270 & 0.1923 & 0.6648 \\\Xhline{1.5pt}
\end{tabular}
\end{table}

\paragraph{\textbf{The Impact of Query Importance Weighting on GWS.}}
%
%
In Table~\ref{table:ablation}, we report the results with and without query importance weighting. We only focus on self-labeling approach, however, our observations generalize to other GWS re-labeling approaches too. According to the table, query importance weigthing using NQC always lead to statistically significant improvements. It helps GWS to focus on more effective examples through weak supervision and query importance weighting is a crucial part of GWS optimization. Future work can explore the impact of various QPP approaches on GWS performance.


\paragraph{\textbf{Additional Analysis}}
For a deeper understanding of GWS performance, in this experiment, we focus on query-level performance differences achieved by GWS. In more detail, we focus on the RoBERTa ranking model training using GWS with self-labeling and plot its performance difference with RoBERTa - WS in Figure \ref{fig:diff}. Due to the smoothness of metrics, we only plot NDCG@10 and MAP for a clear demonstration.


Regarding 0.01 as a bound for a notable amount of change, 46.1\% and 43.9\% of the queries are improved over WS in terms of NDCG@10 and MAP for WikiPassageQA, respectively. Considering the proportion of the degraded queries, 16.8\% and 14.1\%, the cases enhanced by GWS are more than the deteriorated cases. For ANTIQUE, 35\% and 50.5\% of the queries are respectively improved over WS in terms of NDCG@10 and MAP, with 19.5\% and 25\% for deteriorated queries. These plots show that the average improvements obtained by GWS are not dominated by drastic increases in a few queries. 


\section{Conclusion and Future Work}
In this work, we proposed generalized weak supervision (GWS), a generic framework for training retrieval models without requiring any manually labeled training data. Based on weak supervision, which automatically produces training data using existing retrieval models, we generalized the definition of weak labeler to include the weakly supervised models themselves. We provided four implementations of the GWS framework: self-labeling, cross-labeling, joint cross- and self-labeling (JCS), and greedy multi-labeling. We also presented  the theoretical relationship between GWS and the Expectation-Maximization algorithm. Besides, we provided a query importance weighting based on query performance prediction for effective training of GWS models.

In the experiment, we evaluated GWS on two datasets: WikiPassageQA and ANTIQUE. Our experiments showed that GWS achieves substantial improvements compared to  weak supervision in all cases. We observed larger improvements when the power of self and cross labelings are combined (i.e., in JCS and greedy multi-labeling). Furthermore, we showed that query selection via an unsupervised query performance predictor can have significant impact on GWS performance. Our analysis suggested that a large portion of test queries benefit from GWS training.

For future work, we aim to theoretically analyze how GWS affect the training of neural ranking models. Besides, we intend to extend the GWS framework by leveraging multiple weak labelers as well as multiple query performance predictors in order to minimize the noise introduced by the weak labels.

\begin{acks}
This work was supported in part by the Center for Intelligent Information Retrieval, and in part by an Alexa Prize grant from Amazon. Any opinions, findings and conclusions or recommendations expressed in this material are those of the authors and do not necessarily reflect those of the sponsor.
\end{acks}


\bibliographystyle{ACM-Reference-Format}
\bibliography{sample-bibliography}

\end{document}